\documentstyle[epsfig]{aipproc}
\def\be{\begin{equation}}
\def\ee{\end{equation}}
\def\bea{\begin{eqnarray}}
\def\eea{\end{eqnarray}}
\def\CPbar{\hbox{{\rm CP}\hskip-1.80em{/}}}
\def\D0{D\O~}
\def\pbarp{ \bar{{\rm p}} {\rm p} }
\def\pp{ {\rm p} {\rm p} }
\def\ifb{ {\rm fb}^{-1} }
\def\del{\partial }
\def\ra{\rightarrow}
\def\f{\frac}
\def\dis{\displaystyle}
\def\f{\frac}

\def\dis{\displaystyle}
\def\to{\rightarrow}
\def\To{\Rightarrow}
\def\tt{\widetilde{\cal T}}

\renewenvironment{thebibliography}[1]
        {\begin{list}{\arabic{enumi}.}
        {\usecounter{enumi}\setlength{\parsep}{0pt}
\setlength{\leftmargin 0.5cm}{\rightmargin 0pt}
         \setlength{\itemsep}{0pt} \settowidth
        {\labelwidth}{#1.}\sloppy}}{\end{list}}
\begin{document}
\def\be{\begin{equation}}
\def\ee{\end{equation}}
\def\bea{\begin{eqnarray}}
\def\eea{\end{eqnarray}}
\def\CPbar{\hbox{{\rm CP}\hskip-1.80em{/}}}
\def\D0{D\O~}
\def\pbarp{ \bar{{\rm p}} {\rm p} }
\def\pp{ {\rm p} {\rm p} }
\def\ifb{ {\rm fb}^{-1} }
\def\del{\partial }
\def\ra{\rightarrow}

\setcounter{footnote}{0}
\renewcommand{\thefootnote}{\fnsymbol{footnote}}

\title{{\Large {\bf Quartic Gauge Boson Couplings
      }}\hspace*{0.1cm}\thanks{
Invited talk presented at the
Workshop on `` {\it Physics at the First Muon Collider
and at the Front End of a Muon Collider} '', November~6-9, 1997, 
at Fermi National Accelerator Laboratory, Batavia, IL, USA.  
To be published in the conference
proceedings, eds. S.~Geer and R.~Raja. 
}   \\[0.45cm] }

\vspace{0.5cm}
\author{\normalsize  {\sc Hong-Jian He}} 
\address{  
Department of Physics and Astronomy, Michigan State University\\
East Lansing, Michigan 48824, USA\\
( E-mail: hjhe@pa.msu.edu )
}

\maketitle

\vspace*{-6.5cm}
hep-ph/9804210    \hfill  MSUHEP-71120
\vspace*{5.6cm}

{\normalsize 
\begin{abstract}
We review the recent progress in studying the anomalous 
electroweak quartic gauge boson couplings (QGBCs) 
at the LHC and the next generation high energy $e^\pm e^-$ 
linear colliders (LCs). The main focus is put onto the strong 
electroweak symmetry breaking scenario in which the non-decoupling 
guarantees sizable new physics effects for the QGBCs. After commenting 
upon the current low energy indirect bounds and summarizing
the theoretical patterns of QGBCs predicted by the typical 
resonance/non-resonance models, we review our systematic model-independent 
analysis on bounding them via $WW$-fusion and $WWZ/ZZZ$-production. 
The interplay of the two production mechanisms and the important role 
of the beam-polarization at the LCs are emphasized. The same physics may 
be similarly and better studied at a multi-TeV muon collider 
with high  luminosity.\\[0.4cm]
PACS number(s): 11.30.Qc, 11.15.Ex, 12.15.Ji, 14.70.Pw 
\end{abstract}
}

\setcounter{footnote}{0}
\renewcommand{\thefootnote}{\arabic{footnote}}

\vspace{0.8cm}
\section*{\normalsize\bf 1. I\lowercase{ntroduction}}
\vspace{-0.5cm}
\indent\indent

The non-Abelian gauge structure of the standard model (SM)
predicts the presence of electroweak
quartic gauge boson couplings (QGBCs) besides
the couplings of triple gauge bosons. The electroweak symmetry
breaking (EWSB) sector involves 
the would-be Goldstone boson~\cite{GB} dynamics
which generates the longitudinal components for $W^\pm ,Z^0$ so that
they acquire the observed masses. Despite the astonishing success of the
SM at scales up to 
$O(100{\rm GeV})$~\cite{search0,search}, this EWSB sector 
remains unverified~\cite{mike}. Any new physics in the underlying
Goldstone boson dynamics will cause the gauge boson self-interactions
to deviate from the SM. The quartic gauge boson interactions are   
particularly interesting because they can involve four longitudinal 
components which, according to the equivalence theorem~\cite{CG-et,et},
manifest at high energies as pure Goldstone boson interactions 
(that is independent of the SM gauge couplings).
To unambiguously test the couplings of the
quartic gauge boson interactions (QGBCs),  the high energy  
$WW$-fusion and triple gauge boson production processes have to
be used, where the QGBCs directly appear at the tree level.
It is therefore important to study how the future high energy colliders 
(such as the CERN LHC and $e^\pm e^-$ 
linear colliders~\cite{NLC,LC-KEK})
can sensitively probe the QGBCs for unveiling the mystery of 
the EWSB mechanism.

The EWSB sector can interact weakly or strongly. The weakly coupled
case (such as supersymmetric models~\cite{susy}) ensures the new physics
at higher scales to have {\it decoupling} property\cite{decouple}
at low scales, while in the strongly interacting
scenario\cite{peskin-snow}
the nondecoupling guarantees the new physics scale to lie
below or at $4\pi v\sim 3$~TeV~\cite{georgi}. 
In the former case the light Higgs boson(s) plus superpartners have to be
first discovered,
while for the latter we expect sizable new physics deviations 
showing up in the quartic (and triple) gauge boson couplings, which is
the focus of this review.      Below the new physics scale $\Lambda$,
all the new physics effects in the EWSB sector
can be parametrized by a complete set of the next-to-leading order
(NLO) effective operators
of the electroweak chiral Lagrangian (EWCL)~\cite{EWCL}, in which the 
$SU(2)_L \otimes U(1)_Y$ gauge symmetry is nonlinearly realized\footnote{
It is advised that whenever the decoupling theorem~\cite{decouple} becomes 
ineffective, the nonlinear realization should better apply.}.
Without experimental observation on any
new light resonance~\cite{search0,search}, this effective
field theory approach~\cite{weinberg,georgi} 
provides the most economic description of
the possible new physics effects. 
Among the complete set of the fifteen NLO operators, 
five of them characterize only the quartic
gauge interactions~\cite{EWCL}: 
\be
\begin{array}{lc}
\left\{
\begin{array}{lll}
{\cal L}_4 & \, = & \ell_4  \left(\frac{v}{\Lambda}\right)^2
              [{\rm Tr}({\cal V}_{\mu}{\cal V}_\nu )]^2 ~,\\ [0.2cm] 
{\cal L}_5 & \, = & \ell_5 \left(\frac{v}{\Lambda}\right)^2
              [{\rm Tr}({\cal V}_{\mu}{\cal V}^\mu )]^2 ~;
\end{array}\right\}~~
& (~SU(2)_c:~\surd~)  \\[0.65cm]
\left\{
\begin{array}{lll}
{\cal L}_6 & = &\ell_6\left(\frac{v}{\Lambda}\right)^2 
[{\rm Tr}({\cal V}_{\mu}{\cal V}_\nu )]
{\rm Tr}({\cal T}{\cal V}^\mu){\rm Tr}({\cal T}{\cal V}^\nu) ~,\\[0.2cm]
{\cal L}_7 & = & \ell_7\left(\frac{v}{\Lambda}\right)^2 
[{\rm Tr}({\cal V}_\mu{\cal V}^\mu )]
{\rm Tr}({\cal T}{\cal V}_\nu){\rm Tr}({\cal T}{\cal V}^\nu) ~,\\[0.2cm]
{\cal L}_{10} & = &\ell_{10}\left(\frac{v}{\Lambda}\right)^2 
\frac{1}{2}
[{\rm Tr}({\cal T}{\cal V}^\mu){\rm Tr}({\cal T}{\cal V}^{\nu})]^2 ~.
\end{array}\right\}~~
&  (~SU(2)_c:~\times~)
\end{array}
\label{eq:Leff}
\ee 
In (\ref{eq:Leff}), {\small $~{\cal V}_\mu \equiv (D_\mu U)U^\dag~$, 
$D_{\mu}U  =  \partial_{\mu}U + {\bf W}_{\mu}U -U{\bf B}_{\mu}~$,
${\bf W}_{\mu}\equiv ig W^a_{\mu}\tau^a/{2}$,
${\bf B}_{\mu}\equiv ig^{\prime}B_{\mu}\tau^3/{2}$, 
$~U  =  \exp [i\tau^a\pi^a/v ]~$} (with 
$\pi^a$ the would-be Goldstone boson field), and
{\small $~{\cal T}\equiv U\tau_3 U^\dag~$} 
is the custodial $SU(2)_c$-violation operator.
Here, the operators $~{\cal L}_{4,5}~$ conserve $SU(2)_c$
while $~{\cal L}_{6,7,10}~$ violate $SU(2)_c$. The dependence on 
$~v~$ and $~\Lambda~$ is factorized out so that the dimensionless 
coefficient $\ell_n$  of the operator ${\cal L}_n$ is naturally
of $O(1)$~~\cite{georgi}.
Because they contain {\it only} QGBCs
these five operators   
cannot be directly tested via their tree-level contributions at low energies
and are therefore least constrained from the current data. 
So far, only some rough estimates have been made
by inserting them into the one-loop corrections and keeping the $\log$-terms
only\footnote{The ignored constant contributions plus the new loop 
counter-terms are of the same order of magnitude as the $\log$-terms.
So, some uncertainties (like a factor of 2 to 3) 
may naturally exist in these estimates.}. 
Here is a recent estimate at
$90\%$~C.L. by choosing $\Lambda =2$~TeV and setting only one parameter
nonzero at a time~\cite{eboli,global}:
\be
\begin{array}{c}
-4 \leq \ell_4 \leq 20~,~~~~-10 \leq \ell_5 \leq 50 ~;\\
-0.7 \leq \ell_6 \leq 4~, ~~~~-5 \leq \ell_7 \leq 26 ~,~~~~
-0.7 \leq \ell_{10} \leq 3~.
\end{array}
\label{eq:LEPbound}
\ee
(\ref{eq:LEPbound}) shows that the bounds 
on the $SU(2)_c$ symmetric parameters
$\ell_{4,5}$ are about an order of magnitude above their
natural size of $O(1)$; while the allowed range for the $SU(2)_c$-breaking 
parameters $\ell_{6-10}$ is about a factor of $O(10-100)$
larger than that for $\ell_0=\f{\Lambda^2}{2v^2}\Delta\rho
\left(=\f{\Lambda^2}{2v^2}\alpha T\right)$ derived from
the $\rho$ (or $T$) parameter:
$~0.052 \leq \ell_0 \leq 0.12$~\cite{global},
for the same $\Lambda$ and confidence level.
To directly test the EWSB dynamics, it is 
crucial to probe these QGBCs at future 
high energy scattering processes where their contributions can be greatly 
enhanced due to the sensitive power-dependence on 
the scattering energy \cite{global}.

\vspace{0.5cm}
\section*{\normalsize\bf
2. Q\lowercase{uartic} G\lowercase{auge} B\lowercase{oson}
I\lowercase{nteractions} \lowercase{and} 
U\lowercase{nderlying} M\lowercase{odels}  }
\vspace{-0.3cm}
\indent\indent

So far the full theory underlying this effective EWCL 
is not determined, it is thus important to analyze how the typical 
resonance/non-resonance models contribute to these EWSB parameters. 
Knowing the theoretical sizes and patterns of these parameters 
tells how to use the phenomenological bounds derived in following sections
for discriminating different new physics models. We mainly focus
on the quartic gauge boson interactions (\ref{eq:Leff}) and 
consider~\cite{lc-vvv} 
typical models such as a heavy scalar ($S$), a vector ($V^a_\mu$)
and an axial vector ($A^a_\mu$) for the resonance scenario, 
and the new heavy doublet
fermions for the non-resonance scenario. \\[0.15cm]

\noindent
{\bf $\bullet$~A Non-SM Singlet Scalar}~~~~
Up to dimension-$4$
and including both $SU(2)_c$ conserving and breaking effects, we write
down the most general Lagrangian for a singlet scalar which is invariant
under the SM gauge group $SU(2)_L\otimes U(1)_Y$:
\be
\begin{array}{ll}
{\cal L}_{\rm eff}^{S} ~= & 
\dis\f{1}{2}\left[\partial^\mu S\partial_\mu S -M_S^2S^2\right] - V(S)\\
&  -\dis\left[\f{\kappa_s}{2}vS+\f{\kappa_s^{\prime}}{4} S^2\right]
  {\rm Tr}\left[{\cal V}_\mu {\cal V}^\mu \right] 
  -\left[\f{\tilde{\kappa}_s}{2}vS+\f{\tilde{\kappa}_s^{\prime}}{4} S^2\right]
   \left[{\rm Tr} {\cal T}{\cal V}_\mu \right]^2 
\end{array}
\label{eq:scalar}
\ee 
where $V(S)$ only contains Higgs self-interactions. The SM Higgs boson
corresponds to a special parameter choice: 
$~\kappa_s=\kappa_s'=1,~ \tilde{\kappa}_s=\tilde{\kappa}_s'=0~$ and 
{\small $~V(S)=V(S)_{\rm SM}~$.} 
A heavy scalar can be integrated out
from low energy spectrum and 
the corresponding contributions to (\ref{eq:Leff}) are derived as:
\be
\ell_4^s =0~,~~~\ell_5^s =\f{\kappa_s^2}{8} \geq 0~;~~~~~
\ell_6^s =0~,~~~\ell_7^s=\f{\kappa_s\tilde{\kappa}_s}{4}~,~~
\ell_{10}^s =\f{\tilde{\kappa}_s^2}{8} \geq 0 ~.
\label{eq:S-pattern}
\ee
In (\ref{eq:S-pattern}), the deviation from ~$\kappa_s =1$
and $\tilde{\kappa}_s=0$~ signals a {\it non-SM Higgs boson.}\\[0.15cm]

\noindent
{\bf $\bullet$~Vector and Axial-Vector Bosons}~~~~
The $S$-parameter measurement at LEP disfavors the 
naive QCD-like dynamics for the EWSB~\cite{peskin-snow}, 
where the vector $\rho_{\rm TC}$ is the lowest new resonance
in the TeV regime. This suggests a necessity of including the axial-vector
boson\cite{bess} 
in a general formalism for modeling the non-QCD-like dynamics. 
We consider the vector $V^a_\mu$ and axial-vector $A^a_\mu$ fields as the
weak isospin triplets of custodial $SU(2)_c$. 
$\{V,A\}$ transform under the SM global $SU(2)_c$ as {\small 
$~~
\widehat{V}_\mu \To \widehat{V}_\mu^{\prime}
                   =\Sigma_v\widehat{V}_\mu\Sigma_v^\dag ,~~~
\widehat{A}_\mu \To \widehat{A}_\mu^{\prime}
                   =\Sigma_v\widehat{A}_\mu\Sigma_v^\dag ,~
~$  }
where
{\small $~\widehat{V}_\mu \equiv V_\mu^a\tau^a/2, 
         ~\widehat{A}_\mu \equiv A_\mu^a\tau^a/2$,}  and 
{\small $~\Sigma_v\in SU(2)_c~$.}
If $\{V,A\}$  are further regarded as gauge fields of a new
local hidden symmetry group {\small ${\cal H}=SU(2)_L'\otimes SU(2)_R'$} 
(with a discrete left-right parity)~\cite{bess},
we can write down the following general Lagrangian (up to
two derivatives), in the {\it unitary gauge} of the group ${\cal H}$~\footnote{
By ``unitary gauge'' we mean a gauge containing no new Goldstone
boson other than the three ones for generating the longitudinal
components of the {\it known} $W,Z$.
In fact, it is not essentially necessary to introduce such a new local symmetry
${\cal H}$ for $\{V,A\}$~\cite{georgi2} 
since ${\cal H}$ has to be broken anyway and
$\{V,A\}$  can be treated as matter fields~\cite{CCWZ}. The hidden
local symmetry formalism is more restrictive on the allowed free-parameters
($\kappa_n$'s etc) due to the additional assumption 
about that new local group ${\cal H}$.} 
and with both $SU(2)_c$-conserving and -breaking 
effects included\footnote{In the literature~\cite{bess}, only the
$SU(2)_c$-conserving operators were given.},

{\small 
\be
\begin{array}{l}
{\cal L}_{\rm eff}^{VA} =
{\cal L}^{VA}_{\rm kinetic} -v^2\left[
\kappa_0 {\rm Tr}\overline{\cal V}_\mu^2 + 
\kappa_1 {\rm Tr}\left(J^V_\mu -2V_\mu\right)^2+
\kappa_2 {\rm Tr}\left(J^A_\mu+2A_\mu\right)^2+
\kappa_3 {\rm Tr}A_\mu^2\right. \\[0.25cm]
~~~~~~\left.
+\tilde{\kappa}_0\left[{\rm Tr}\tt \overline{\cal V}_\mu\right]^2
+\tilde{\kappa}_1\left[{\rm Tr}\tt (J^V_\mu -2V_\mu )\right]^2
+\tilde{\kappa}_2\left[{\rm Tr}\tt (J^A_\mu +2A_\mu )\right]^2
+\tilde{\kappa}_3\left[{\rm Tr}\tt A\right]^2 \right]
\end{array}
\label{eq:VA}
\ee
}
where
\vspace{-0.2cm}
$$
\begin{array}{ll}
\left\{\begin{array}{l} J^V_\mu =J^L_\mu +J^R_\mu \\[0.2cm] 
                        J^A_\mu =J^L_\mu -J^R_\mu 
       \end{array} \right.
&~~~~~
\left\{\begin{array}{l}
J^L_\mu =\xi^\dag D^L_\mu\xi 
        =\xi^\dag\left(\partial_\mu\xi + W_\mu\xi\right)\\[0.2cm]
J^R_\mu =\xi D^R_\mu\xi^\dag 
        =\xi\left(\partial_\mu\xi^\dag +B_\mu\xi^\dag\right)
\end{array}  \right.
\end{array} 
$$
and, by definition,
{\small $~V_\mu\equiv i\tilde{g}\widehat{V}_\mu =i\tilde{g}V^a_\mu\tau^a/2,
         ~A_\mu\equiv i\tilde{g}\widehat{A}_\mu =i\tilde{g}A^a_\mu\tau^a/2$,}
{\small  $\overline{\cal V}_\mu \equiv U^\dag D_\mu U 
=U^\dag{\cal V}_\mu U$,} {\small $\tt =\tau^3 =U^\dag{\cal T}U$,}
and {\small $U \equiv \xi^2$.}    
Here $\tilde{g}$ is the gauge coupling of the group ${\cal H}$.
Among the above four new $SU(2)_c$-conserving 
parameters $\kappa_n$'s, $\kappa_0$ is determined
by normalizing the Goldstone kinematic term: 
$~\kappa_0 =-4\kappa_2\kappa_3/(4\kappa_2+\kappa_3)~$. 
After eliminating the $V$ and $A$ fields in the heavy mass expansion,
we derive  $\ell_{n}$'s  below:
\be
{\small
\begin{array}{ll}
\left\{\begin{array}{l}  \ell_4=\ell_4^v+\ell_4^a \\[0.3cm]
                    \ell_5=\ell_5^v+\ell_5^a \\[0.3cm]
                    \ell_6=\ell_6^v+\ell_6^a \\[0.3cm]
                    \ell_7=\ell_7^v+\ell_7^a \\[0.3cm]
                    \ell_{10}=\ell_{10}^v+\ell_{10}^a    
\end{array}\right.
&~~~~~~
\left\{\begin{array}{l}
    \ell_4^v=-\ell_5^v=1/[2\sqrt{2}\tilde{g}v\Lambda^{-1}]^2>0 \\[0.3cm]
    \ell_4^a=-\ell_5^a=\left[\eta^2(\eta^2-2)+16\tilde{\eta}^2\right]/
               [2\sqrt{2}\tilde{g}v\Lambda^{-1}]^2 \\[0.3cm]
    \ell_6^{v}=\ell_7^{v}=0 \\[0.3cm]
    \ell_6^{a}=-\ell_7^{a}
     =-\tilde{\eta}\left[4(3-\eta^2)\tilde{\eta}+(1-\eta^2)\eta\right]/
[2\sqrt{2}\tilde{g}v\Lambda^{-1}]^2 \\[0.3cm]
    \ell_{10}^{v}=\ell_{10}^{a}=0
   \end{array}\right.
\end{array}
\label{eq:VA-pattern}
}
\ee
in which 
\vspace*{-0.3cm}
\be 
{\small
~\eta = \dis\f{4\kappa_2}{4\kappa_2+\kappa_3}~,~~~~
\tilde{\eta} =\dis\f{2\kappa_2+4\tilde{\kappa}_2}
{\left(4\kappa_2+\kappa_3\right)+2\left(4\tilde{\kappa}_2+
\tilde{\kappa}_3\right)} - \f{2\kappa_2}{4\kappa_2+\kappa_3}~,
\label{eq:VA-para}
}
\ee
and $~\Lambda =\min\{M_V,M_A\}~$. 
After ignoring the SM gauge couplings $g$ and $g'$,   
{\small $\{M_V,M_A\}\simeq \left\{\tilde{g}v\sqrt{\kappa_1},~
\tilde{g}v\sqrt{\kappa_2+\kappa_3/4} \right\}$, }
at the leading order.
In (\ref{eq:VA-pattern}), the factor
{\small $~1/[\tilde{g}v\Lambda^{-1}]^2\simeq \kappa_1 (\Lambda /M_V)^2
=O(\kappa_1)$~} and all $SU(2)_c$-breaking terms depend on $\tilde{\eta}$~.
Note that the $SU(2)_c$-symmetric contribution 
from the axial-vector boson interactions to {\small ~$\ell_4^a=-\ell_5^a$~ }
becomes negative for {\small $~|\eta | <\sqrt{2}~$}, 
while the summed contribution {\small $~\ell_4=-\ell_5
=\left[ (\eta^2-1)^2+16\tilde{\eta}^2 \right]/
[2\sqrt{2}\tilde{g}v\Lambda^{-1}]^2\geq 0~$.}
The deviation of $\eta$ and/or $\tilde{\eta}$ 
from $\eta (\tilde{\eta}) =0$ represents the {\it non-QCD-like}
EWSB dynamics. \\[0.15cm]

\noindent
{\bf $\bullet$~Heavy Doublet Fermions}~~~~
Take for instance a 
model of one flavor heavy chiral fermions which form a left-handed weak 
doublet $(U_L,D_L)^T$ and right-handed singlets $\{U_R, D_R\}$, and joins
a new strong $SU(N)$ gauge group in its fundamental representation.
 Their small mass-splitting
breaks the $SU(2)_c$ and is characterized by the parameter
$~\omega =1-\left[M_U/M_D\right]^2~$. The anomaly-cancellation is ensured
by assigning the $\{U,D\}$ electric charges as 
~$\{+\f{1}{2},-\f{1}{2}\}$~.
By taking $\{U, D\}$ as the source of the EWSB,
the $W,Z$ masses can be generated by heavy fermion loops. The new contributions
to the quartic gauge couplings of $W/Z$ come from the {\it non-resonant}
$\{U, D\}$ box-diagrams.
 The leading results in the $1/M_{U,D}$ and $\omega$ expansions 
are summarized below:
\be
\dis\ell_4^f=-2\ell_5^f=\left[\f{\Lambda}{4\pi v}\right]^2\f{N}{12}>0~;~~~
\dis\ell_6^f=-\ell_7^f=
-\left[\f{\Lambda}{4\pi v}\right]^2\f{7N}{240}\omega^2~,~~~
\dis\ell_{10}=0~;
\label{eq:fermion-pattern}
\ee
in which $~\Lambda =\min\{M_U,M_D\}~$.

\vspace{0.5cm}
\section*{\normalsize\bf
3. A G\lowercase{lobal} A\lowercase{nalysis} \lowercase{on} 
P\lowercase{robing} QGBC\lowercase{s} 
\lowercase{versus} TGBC\lowercase{s} \lowercase{at the} LHC}  
\vspace{-0.3cm}
\indent\indent

The general EWCL formalism \cite{EWCL} contains in total
15 NLO new operators whose coefficients ($\ell_n$'s) depend on 
the details of the underlying dynamics 
as exemplified in the previous section.  
It is shown\cite{global} that, 
except for $\ell_{0,1,8}$ ($S,T,U$), 
the current data only bound a few triple gauge boson couplings
(TGBCs) to $O(10)$ at the $1\sigma$-level
and give no direct tree-level bound on QGBCs. The rough estimates of the
bounds from 1-loop corrections still allow QGBCs to be of $O(5-50)$.
For a {\it complete} test of the EWSB sector in discriminating different 
dynamical models, all these TGBCs and QGBCs ($\ell_n$'s)  
have to be measured through various high energy $VV$-fusion and 
$f\bar{f}^{(\prime )}$-annihilation processes. ($V^a=W^\pm ,Z^0$.)
For this purpose, a systematic global analysis~\cite{global}
has been carried out which reveals the important overall physical
pictures and guides us for further elaborate precise numerical studies
(cf. Secs.~4-5). In performing such a global analysis we developed a precise
electroweak power counting rule (\`{a} la Weinberg)
for conveniently estimating {\it all} high energy scattering amplitudes 
and formulated the equivalence theorem (ET)~\cite{et} as a {\it necessary} 
physical criterion for sensitively probing the EWSB dynamics. 
Applying this counting method, 
we have carried out a systematic analysis for
all $~V^aV^b \ra V^cV^d~$ and 
$~f\bar{f}^{(\prime )}\ra V^aV^b,V^aV^bV^c~$ processes
by estimating the contributions to their $S$-matrix elements
from both the leading order operator 
up to one-loop and the other 15 NLO operators at the tree-level. 
Based upon the basic features of the chiral perturbation expansion, we
further build the following electroweak power counting hierarchy for
the $S$-matrix elements,\footnote{For 
$~f\bar{f}^{(\prime )}\ra VVV~$ amplitudes, there is an additional factor
$1/f_\pi$ by dimentional counting.}
\be
\dis
\frac{E^2}{f_\pi^2}\gg 
\left[\frac{E^2}{f_\pi^2}\frac{E^2}{\Lambda^2},~g\frac{E}{f_\pi}\right] \gg 
\left[g\frac{E}{f_\pi}\frac{E^2}{\Lambda^2}, ~g^2\right] \gg 
\left[g^2\frac{E^2}{\Lambda^2}, ~g^3\frac{f_\pi}{E}\right] \gg 
\left[g^3\frac{Ef_\pi}{\Lambda^2},~g^4\frac{f^2_\pi}{E^2}\right]\gg 
g^4\frac{f_\pi^2}{\Lambda^2}~~.
\label{eq:hierarchy}
\ee
In the typical TeV region, for 
$~~E\in (750\,{\rm GeV},~1.5\,{\rm TeV})$, this gives:
$$
\begin{array}{c}
(9.3,37)\gg \left[(0.55,8.8),(2.0,4.0)\right]\gg 
\left[ (0.12,0.93),(0.42,0.42)\right] \gg \\
\left[ (0.025,0.099),(0.089,0.045)\right]\gg 
\left[ (5.3,10.5),(19.0,4.7)\right]\times 10^{-3}\gg 
(1.1,1.1)\times 10^{-3} ~,
\end{array}
$$
where $E$ is taken to be the invariant mass of the $VV$ pair.
This power counting hierarchy can be nicely understood. 
In (\ref{eq:hierarchy}), from left to right, the hierarchy
is built up by increasing either the number of derivatives (i.e. 
power of $E/\Lambda$) or the number of external transverse gauge boson 
$V_T$'s (i.e. the power of gauge couplings).  This power counting
hierarchy provides us a theoretical base to classify all
the relevant scattering amplitudes 
in terms of the three essential parameters $E$, $f_\pi$ and $\Lambda$ 
plus possible gauge/Yukawa coupling constants.

At the event-rate-level, we have adopted the usual leading-$\log$
effective vector boson method~\cite{ewa} 
to reasonably and conveniently estimate the $VV$-luminosities. 
In Fig.~1, the rate $|R_B|$ denotes an
intrinsic background defined via the formulation of the ET as a necessary
criterion for the sensitivities to the EWSB \cite{et,global}.
Fig.~1 shows that, at the $14$TeV LHC with $\int{\cal L}=100$fb$^{-1}$
Luminosity and for $\Lambda =2$TeV, the $W^+W^+$-fusion is most sensitive to 
$\ell_{4,5}$ (QGBCs) and marginally sensitive to $\ell_{3,9,11,12}$; 
while the $q\bar{q}'\ra W^+Z$ annihilation can best probe $\ell_{3,11,12}$ and
marginally test $\ell_{8,9,14}$.  
Hence, the $VV$-fusions and $f\bar{f}^{(\prime )}$-annihilations are 
{\it complementary} in probing the different sets of these NLO parameters 
(the QGBCs and TGBCs) at the LHC.

\vspace{0.5cm}
\section*{\normalsize\bf
4. M\lowercase{easuring} \lowercase{the} 
QGBC\lowercase{s}  \lowercase{via} $WW$-F\lowercase{usion} 
P\lowercase{rocesses}}
\vspace{-0.3cm}
\indent\indent

Though the LHC will give the first direct test on these new quartic gauge
boson couplings (QGBCs), the large backgrounds limit its sensitivity
and cutting off the backgrounds significantly reduces the event rate.
As shown in Ref.~\cite{B-Y},
for the non-resonance $W^\pm W^\pm$ production channels in the TeV regime
only around $10$ signal events were predicted at the LHC 
with a $100$~fb$^{-1}$ annual luminosity after
imposing necessary cuts in the gold-plated modes (by pure leptonic
decays). The corresponding study at the TeV $e^\pm e^-$ LCs opens a more
exciting possibility~\cite{kek,han-lc}.  

In this and next sections we review how 
to make further precision constraints for the QGBCs
via the $WW$-fusion~\cite{lc-he,lc-vv},
$WWZ/ZZZ$-production~\cite{lc-vvv}\footnote{
The $WWZ/ZZZ$-production in the SM was first studied 
in Ref.~\cite{vvv-han}, and later some analyses on including the 
anomalous couplings have also appeared~\cite{vvv-other} for the
case of unpolarized $e^\mp$ beams. For a very recent study similar
to Ref.~\cite{lc-vvv} for $WWZ/ZZZ$-production, see Ref.~\cite{eboli}.}, 
and their interplay at LCs~\cite{lc-vvv,vvv-BSMV}, 
which is much cleaner than the LHC so that the final state 
$W/Z$'s can be detected via the dijet mode and with large branching ratios.
Due to the limited calorimeter energy resolution, the misidentification
probability of $W$ versus $Z$ and the rejection of certain fraction of
diboson events should be considered~\cite{han-lc}. 
Inclusion of the leptonic
decay of $Z$ to $e^-e^+$ and $\mu^-\mu^+$ is also useful. 
According to the study of Ref.~\cite{han-lc}, the (mis)identification
probabilities of $W$ and $Z$ via jet-decay mode can be derived as
\be
\begin{array}{rcl}
W & \To & 85\%~W,~~10\%~Z,~~5\%~{\rm reject}~,\\[0.15cm]
Z & \To & 22\%~W,~~74\%~Z,~~4\%~{\rm reject}~.\\
\end{array}
\label{eq:misiden}
\ee
The detection
efficiencies for $WW$, $ZZ$ and $WZ$ final states are thus estimated below,
which are about $34\%$:
\be
\begin{array}{l}
 \epsilon_{WW}=[0.68\times 0.85]^2  = 33.4\%~,~~~~~
 \epsilon_{ZZ}=[0.70\times 0.74 +0.067 ]^2 = 34.2\%~,\\[0.15cm]
 \epsilon_{WZ}=[0.68\times 0.85][0.70\times 0.74 +0.067 ]
              =33.8\%~.
\end{array}
\label{eq:efficiency}
\ee

To completely determine
all the QGBCs, we need at least five independent processes. 
From $WW$-fusions alone, we can have  
\be
\begin{array}{lll}
{\rm Full~process:} &  {\rm Sub-process:} & {\rm Relevant~parameter:}\\[0.3cm]
e^-e^+ \to \nu\bar{\nu}W^-W^+ ~, & (W^-W^+\to W^-W^+),~
& (\ell_{4,5})~, \\[0.15cm]
e^-e^- \to \nu\bar{\nu}W^-W^- ~, & (W^-W^-\to W^-W^-),
& (\ell_{4,5})~; \\[0.25cm]
e^-e^+ \to \nu\bar{\nu}ZZ ~,     & (W^-W^+\to ZZ),
& (\ell_{4,5};~\ell_{6,7})~, \\[0.15cm]
e^-e^+ \to e^\pm\nu W^\mp Z ~,~~& (W^\mp Z\to W^\mp Z ),
& (\ell_{4,5};~\ell_{6,7})~,  \\[0.15cm]
e^-e^+ \to e^-e^+ ZZ ~,~~& (ZZ\to ZZ),
& ([\ell_4+\ell_5]+2[\ell_6+\ell_7+\ell_{10}])~; 
\end{array}
\label{eq:fusion}
\ee
where in the round backets the corresponding 
fusion (signal) sub-processes are given.
We see that $\ell_{4,5}$ can be cleanly tested via the first two
processes in (\ref{eq:fusion}), as shown by Fig.~2.
(All our plots have chosen the new physics cutoff as $~\Lambda =2$~TeV
and the numerical results for other values of
$~\Lambda~$ can be obtained via re-scaling.)
By including the third and fourth reactions
$\ell_{6,7}$ can be further disentangled. Finally the 
fifth channel provides a unique probe on $\ell_{10}$. Though this scheme
is complete in principle, the realistic situation is more involved. 
Note that the rate of the last reaction in (\ref{eq:fusion}) 
is significantly lower than all others due to the double suppressions 
of the $e$-$e$-$Z$ couplings
while the fourth channel has huge backgrounds which are uneasy to 
overcome~\cite{han-lc,lc-vv}. 
But $ZZ\to ZZ$ also has an advantage due to 
the absence fusion-type backgrounds and the triple gauge boson
couplings have no contribution either. This makes it relatively cleaner
than others.  Since the parameter $\ell_{10}$ appears
only in $4Z$ vertex, the above last channel has to be used anyway when
only the fusion mechanism is studied. 
(For the process $e^-e^+\to ZZZ$ on $\ell_{10}$, see Sec.~5.)
Since the large backgrounds make the $WZ$-channel
less useful (see Fig.~4a below), we propose to use the  
production $e^-e^+\to WWZ$ (cf.~Sec.~5) 
to complete this five parameter determination.

From the above analysis, we finally summarize below
the $90\%$~C.L. (one-parameter) fusion-bounds for $\Lambda =2$~TeV
at a later stage of the LC with the energy $\sqrt{s}=1.6$~TeV
and the integrated luminosity $\int{\cal L}=200$~fb$^{-1}$:
\be
{\small 
\begin{array}{c}
-0.13 \leq \ell_4 \leq 0.10~,~~~~-0.08 \leq \ell_5 \leq 0.06 ~;\\
-0.22 \leq \ell_6 \leq 0.22~, ~~~~-0.12 \leq \ell_7 \leq 0.10 ~,~~~~
-0.21 \leq \ell_{10} \leq 0.21 ~;
\end{array}
\label{eq:fusion-bound}
}
\ee
which are very stringent. Here we have used
a $90\%$ ($65\%$) polarization for the $e^-$($e^+$) beam.

\vspace{0.5cm}
\section*{\normalsize\bf
5. $WWZ/ZZZ$-P\lowercase{roduction} \lowercase{and its}
I\lowercase{nterplay} \lowercase{with} $WW$-F\lowercase{usion} }
\vspace{-0.3cm}
\indent\indent

To probe the QGBCs (\ref{eq:Leff}), 
we know~\cite{global} that the $WW$-fusion amplitudes have the
highest $E$-power dependence in the TeV regime while the $s$-channel
signals of the $WWZ/ZZZ$-production lose 
an enhancement factor of $(E/v)^2$ relative   
to that of the fusion processes. When the collider energy is reduced by half
(from $1.6$~TeV down to $800$~GeV),     
the sensitivity of the $WW$-fusion decreases by about a factor of $20$
or more~\cite{lc-he,lc-vv}. We thus expect that $ee\to WWZ,ZZZ$
become more important at the earlier phase of the LCs and will be 
competitive with and complementary to the fusion processes 
for the later stages of the
LCs around  $0.8\sim 1$~TeV~\cite{lc-vvv}. In fact, it was revealed
that even at the $1.5/1.6$~TeV, $e^+e^-\to WWZ$ plays a crucial role
in achieving a clean five-parameter 
analysis \cite{lc-vvv,vvv-BSMV}.

To avoid the potential fusion backgrounds from $e^-e^+\to eeZZ,eeWW$, 
we now only add the 
$Z\to \mu^-\mu^+$ decay besides the dijet-decay mode. 
The detection efficiencies for $ZZZ$ and $WWZ$ final 
states are thus estimated to be about $16.8\%$ and $18.4\%$, respectively.  
It turns out that $e^-e^+\to WWZ$ has
huge backgrounds due to the $t$-channel $\nu_e$ or $e$-$\nu_e$ exchange, and 
the kinematic cuts alone help very little. However, we find that such type of 
backgrounds involve the left-handed $W$-$e$-$\nu$ coupling and thus
can be effectively suppressed by using the right(left)-hand polarized 
$e^-(e^+)$ beam. The highest sensitivity is reached by maximally polarizing
{\it both} $e^-$ and $e^+$ beams. 
The crucial roles of the beam polarization and the higher
collider energy for the $WWZ$-production are
demonstrated in Fig.~3a, where $\pm 1\sigma$ exclusion
contours for $\ell_4$-$\ell_5$ are displayed at $\sqrt{s}=0.5,~ 0.8,~ 
1.0$ and $1.6$~TeV,  respectively.
The beam polarization has much less impact on the $ZZZ$ mode, due to  
the almost axial-vector type $e$-$Z$-$e$ coupling. Including the same
polarizations as in the case of the $WWZ$ mode, we find about 
$10-20\%$ improvements on the bounds from the $ZZZ$-production.
Assuming the two beam polarizations ($90\% ~e^-$ and $65\% ~e^+$),
we summarize the final $\pm 1\sigma$ bounds
for both $ZZZ$ and $WWZ$ channels and their combined $90\%$~C.L. contours
for $0.5$~TeV with $\int {\cal L}=50$~fb$^{-1}$ in Fig.~3b
(representing the {\it first direct probe} at the
LC) and for $1.6$~TeV with $\int {\cal L}=200$~fb$^{-1}$ in Fig.~3c 
(representing the {\it best} sensitivity gained
from the final stage of the LC with energy around $1.5/1.6$~TeV).
Note that, the $90\%$~C.L. level bounds on $\ell_4$-$\ell_5$
at 0.5~TeV are within $O(10-20)$, while at 1.6~TeV they sensitively reach 
$O(1)$. The $WWZ$ channel gives the same bounds for $\ell_4$-$\ell_5$ and
$\ell_6$-$\ell_7$, while the $ZZZ$ channel imposes stronger bound
on $\ell_6$-$\ell_7$ due to a factor of 2 enhancement from the $4Z$-vertex.  
$\ell_{10}$ only contributes to $ZZZ$ final state and can be
probed at the similar level.

For comparison, a parallel analysis to Fig.~3b-c is further performed  
for the case without $e^+$-beam polarization
(but with $e^-$ polarization the same as before). For a two-parameter 
($\ell_{4,5}$) study, the results are listed below at $90\%$~C.L.:
\be
{\small 
\begin{array}{lcc}
{\rm at~0.5~TeV:}~~~ & 
-12~(-18)\leq \ell_4 \leq 21~(27), ~&~ -17~(-22)\leq \ell_5 \leq 9.5~(15);
\\[0.18cm]
{\rm at~1.6~TeV:}~~~ & 
-0.50~(-0.67)\leq \ell_4 \leq 1.5~(1.7), 
~&~ -1.3~(-1.5)\leq \ell_5 \leq 0.36~(0.58);
\end{array}
}
\label{eq:e+pol-compare}
\ee
where the numbers in the parentheses denote the bounds from polarizing the
$e^-$-beam alone.  The comparison in (\ref{eq:e+pol-compare}) 
shows that without $e^+$-beam polarization, 
the sensitivity will decrease by about $15\%-60\%$. 
Therefore, making use of the possible $e^+$-beam polarization with a degree 
around $65\%$ is clearly helpful.
In the above, the total rates are used to derive the numerical bounds.
We have further studied the possible improvements by including different
characteristic distributions, 
but no significant increase of the sensitivity is found.

Now, we are ready to analyze the interplay with $WW$-fusion processes.
As noted in Sec.~4,  the $WZ$-channel in (\ref{eq:fusion}) has 
large $\gamma$-induced $eeWW$ background in which one $e$ is lost in the
beam-pipe and one $W$ misidentified as $Z$. A cut on the missing 
$p_\perp (\nu )$ is imposed to specially suppress this background.
Even though, the final sensitivity still turns out to be 
less useful in constraining the $\ell_6$-$\ell_7$ plane
(cf. Fig.~4a)~\cite{lc-vv}. 
To sensitively bound $\{\ell_6,\ell_7\}$ (especially $\ell_6$) well 
below $O(1)$, we propose to use the production $e^-e^+\to WWZ$. 
Fig.~4a demonstrates the interplay of $WW$-fusion and $WWZ$-production for
discriminating the $SU(2)_c$-breaking QGBCs $\ell_{6}$-$\ell_7$
at $\sqrt{s}=1.6$~TeV. 
The $ZZZ$-production can also bound $\ell_{10}$,
in addition to the $eeZZ$ fusion-channel in (\ref{eq:fusion}).
Assuming that $\ell_{4,5;6,7}$ are constrained by the processes mentioned
above, we set their values to the reference point (zero) for simplicity 
and define the statistic significance
$~S=|{\cal N}-{\cal N}_{0}|/\sqrt{{\cal N}_{0}}~$ which is
a function of $\ell_{10}$. Here ${\cal N}$ is the
total event-number while ${\cal N}_0$ is the number at $\ell_{10}=0$.
As shown in Fig.~4b, at 1.6~TeV, the sensitivity of $e^-e^+\to eeZZ$ 
for probing $\ell_{10}$ is better than that of $e^-e^+\to ZZZ$.

In summary,
the first direct probe on these QGBCs will come from the early 
phase of the LC at 500~GeV, where
the $WW$-fusion processes are not useful.
The two mechanisms become more competitive and
complementary at energies $\sqrt{s}\sim 0.8-1$~TeV.
From (\ref{eq:fusion-bound}) and Table~1, we see that
at a later stage of the LC with $\sqrt{s}=1.6$~TeV, the $90\%$~C.L.
one-parameter bounds on $\ell_{4,5}$ from $WWZ/ZZZ$-modes 
are about a factor of $3\sim 6$ weaker than that from $WW$-fusions;
while the bounds on $\ell_{6,7,10}$ are comparable.
In a complete multi-parameter analysis, the $WWZ$-channel is crucial for
determining $\ell_6$-$\ell_7$ even at a 1.6~TeV LC (cf. Fig.~4a).\\[0.2cm]

%
{\small Table~1: Combined $90\%$~C.L. bounds on $\ell_{4-10}$
from $WWZ/ZZZ$-production. For simplicity, we
set one parameter to be nonzero at a time. The bound on $\ell_{10}$ 
comes from $ZZZ$-channel alone.\\[-0.7cm]
} 
\begin{center}
\tabcolsep 1pt
{\small
\begin{tabular}{c||c|c|c|c}
\hline\hline
&&&&\\[-0.2cm]
$\sqrt{s}$~(TeV) & 0.5 & 0.8 & 1.0 & 1.6 \\[0.15cm]
\hline
&&&&\\ [-0.36cm]
$\int{\cal L}$~(fb$^{-1}$) & 50
& 100 & 100 & 200  \\ [0.2cm]
\hline\hline
&&&&\\[-0.33cm]
& $~-9.5\leq\ell_4\leq 11.7~$   & $~-2.7\leq\ell_4\leq 3.2~$
& $~-1.7\leq\ell_4\leq 2.0~$   & $~-0.50\leq\ell_4\leq 0.58~$ \\[0.2cm]
$WWZ/ZZZ$~
& $-9.8\leq\ell_5\leq ~8.9$   & $-3.1\leq\ell_5\leq 2.3$
& $-1.9\leq\ell_5\leq 1.4$   & $-0.54\leq\ell_5\leq 0.36$ \\[0.2cm]
Bounds
& $-5.0\leq\ell_6\leq 5.8$          & $-1.5\leq\ell_6\leq 1.6$
& $-0.95\leq\ell_6\leq 1.0$          & $-0.28\leq\ell_6\leq 0.28$  
\\[0.2cm]
(at 90$\%$C.L.)~
& $-5.0\leq\ell_7\leq 5.7$          & $-1.5\leq\ell_7\leq 1.5$
& $-0.95\leq\ell_7\leq 0.92$          & $-0.28\leq\ell_7\leq 0.26$\\[0.2cm]
& $-4.3\leq\ell_{10}\leq 5.2$       & $-1.4\leq\ell_{10}\leq 1.4$
& $-0.83\leq\ell_{10}\leq 0.88$        & $-0.26\leq\ell_{10}\leq  
0.26$ \\[0.2cm]
\hline
&&&&\\ [-0.36cm]
Range of $|\ell_n|$
& $\leq O(4\sim 10)$    &   $\leq O(1\sim 3)$
& $\leq O(0.8\sim 2)$  &   $\leq O(0.3\sim 0.6)$\\[0.3cm] 
\hline\hline
\end{tabular}
}
\end{center}
\vspace*{0.2cm}

\vspace*{0.3cm}
\section*{\normalsize\bf
6. C\lowercase{oncluding} R\lowercase{emarks} }
\vspace*{0.3cm}

Despite the constantly increasing evidence in supporting
the Standard Model (SM) over the past 30 years, 
we particle physicists have been 
struggling in search for {\it New Physics
Beyond the SM} so far~\cite{search0,search}. Among the numerous ways for
going beyond the SM, the Higgs boson hypothesis~\cite{higgs} stands out. 
The updated direct Higgs search at LEP~\cite{search0} 
puts a $95\%$C.L. lower bound $~m_H \geq 89.3$~GeV. 
Due to the discrepancy between the
precision $Z$-decay asymmetry measurement and the direct Higgs search 
limit, the combined $95\%$C.L. upper Higgs mass bound from the global fit has
been shown to significantly increase toward the TeV regime~\cite{mike}.
However, the unitarity~\cite{uni} 
and triviality~\cite{trivial} theoretically
forbid the SM Higgs mass to go beyond the
TeV scale, at which we are facing an exciting strong 
electroweak symmetry breaking (EWSB) dynamics.
Below the new heavy resonance, 
we have to first probe the EWSB parameters
formulated by means of the electroweak chiral Lagrangian (EWCL), 
among which the quartic gauge boson interactions 
penetrate the pure Goldstone dynamics. 
After commenting upon the low energy indirect bounds and
analyzing the different patterns of these quartic couplings
predicted by the typical resonance/non-resonance models,
we estimate the sensitivity of the LHC to probing these
couplings, and then analyze the constraints on them via
$WWZ/ZZZ$-production and $WW$-fusion   
at the next generation $e^\pm e^-$ linear colliders (LCs). 
The interplay of the two
production mechanisms and the important role of the beam-polarization
at the LCs are revealed and stressed. 

Finally, we remark that
the same physics may be similarly
and better studied at a multi-TeV muon collider (MTMC)
($\sqrt{s}\simeq 3-4$~TeV) with high luminosity 
($\sim 500-1000$~fb$^{-1}$/year) \cite{MC,Gunion}. 
Due to the higher center mass energy of the MTMC, 
certain unitarization on
the EWCL is needed for studying the $WW$-fusions.
The other muon collider options, like
$\mu^-\mu^-$ and $\mu^+\mu^+$ are likely to be as easily achieved as
$\mu^-\mu^+$ mode. Furthermore, the large muon mass relative to the
electron mass makes the initial state photon-radiation of the
muon collider much less severe
than that of the electron collider. 
The two drawbacks of a muon collider in comparison with an electron linear
collider are \cite{Gunion}: (i) substantial beam
polarization ($\geq 50\%$) can be achieved only with a significant
sacrifice in luminosity; (ii) the $\gamma\gamma$ and $\mu\gamma$ options
are probably not very feasible.

\vspace{0.5cm}
\noindent
{\normalsize\bf Acknowledgments}~~~  \\[0.25cm]
I am grateful to the working group convener Tao~Han for invitation
and FermiLab for hospitality.
Special thanks go to Tao~Han and C.--P.~Yuan for carefully reading
the draft and providing many useful suggestions.
I thank them and all other collaborators,
E.~Boos, W.~Kilian, Y.-P.~Kuang, A.~Pukhov and 
P.M.~Zerwas, for productive collaborations \cite{global,lc-vv,lc-vvv}
upon which this review is based.
I am also indebted to K.~Floettmann and R.~Frey for discussing
the $e^\mp$-beam polarizations, and many other colleagues 
such as M.S.~Chanowitz, R.~Casalbuoni, D.~Dominici, K.~Hagiwara,
K.~Hikasa, G.~Jikia, I.~Kuss, A.~Likhoded, C.R.~Schmidt and G.~Valencia 
for useful conversations on this subject. 
This work is supported by the U.S. Natural Science Foundation.

\newpage
\vspace{0.3cm}
\section*{\normalsize\bf  R\lowercase{eferences}}
\vspace{-0.1cm}

%
%
\vspace*{-0.5cm}
\null \vspace{-0.7cm}       
\begin{figure}[T]    
\begin{center}
\vspace{-1.5cm}
\centerline{\epsfig{file=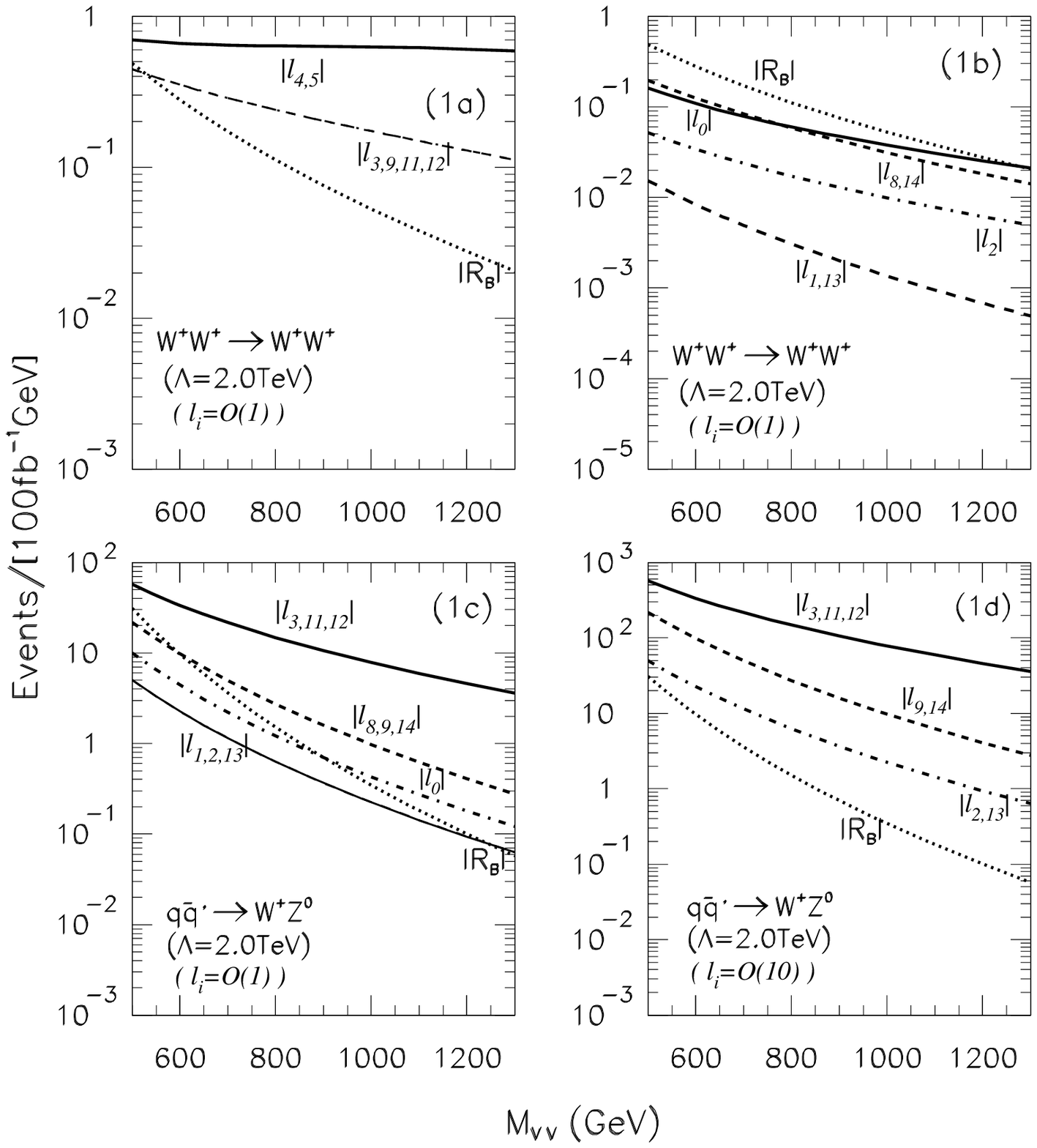,height=21cm,width=17.5cm}}
\end{center}
\vspace{-1.5cm}
\hspace{-0.9cm}
\caption{A classification of the contributions 
from all 15 next-to-leading order operators 
at the $14$~TeV LHC (with $100$~fb$^{-1}$ annual luminosity) 
for $\Lambda = 2$~TeV.}
\label{fig:fig-lhc}
\end{figure}

\begin{figure}[T]   
\begin{center}
\hspace{-2.8cm}
\vspace{-0.4cm}
\epsfig{file=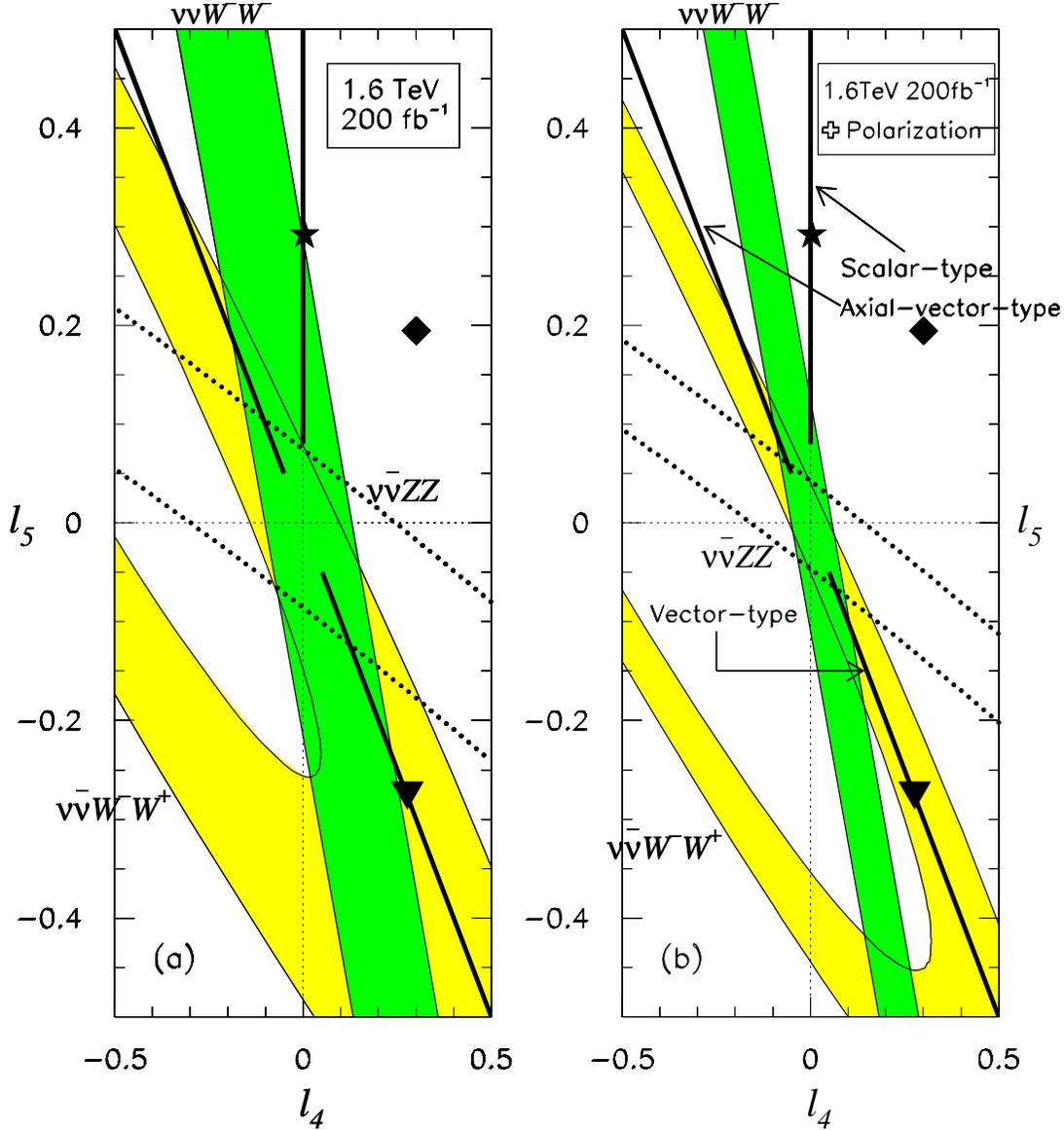,height=18cm,width=17cm}
\end{center}
\vspace{-1cm}
\caption{
Determining the $SU(2)_c$-symmetric parameters $\ell_4$-$\ell_5$ 
at 1.6~TeV $e^-e^+/e^-e^-$ LCs. Here the $\pm1\sigma$ exclusion 
contours are displayed.
(a).~unpolarized case; 
(b).~the case with 90$\%($65$\%)$ polarized $e^-(e^+)$~beam.
Contributions from three types of resonance models (scalar, vector
and axial-vector) to $(\ell_4,\ell_5)$ are shown by the thick solid
lines. The different points on these solid lines correspond to different
values of their couplings to the weak gauge bosons.
Note that for axial-vector-type, it is also possible to have
$\ell_4+\ell_5=0$ with $\ell_4 \geq 0$, i.e., similar to the vector-type
case. This makes the discrimination more involved.
Big-star: from a scalar; 
black-triangle-down: from a vector;
black-lozenge: from mixed contributions of a heavy scalar and vector.
(Here we typically set these heavy resonances around 2~TeV.)
}
\label{fig:fig-L45}
\end{figure}

\begin{figure}[T]   
\vspace{-0.4cm}
\centerline{\epsfig{file=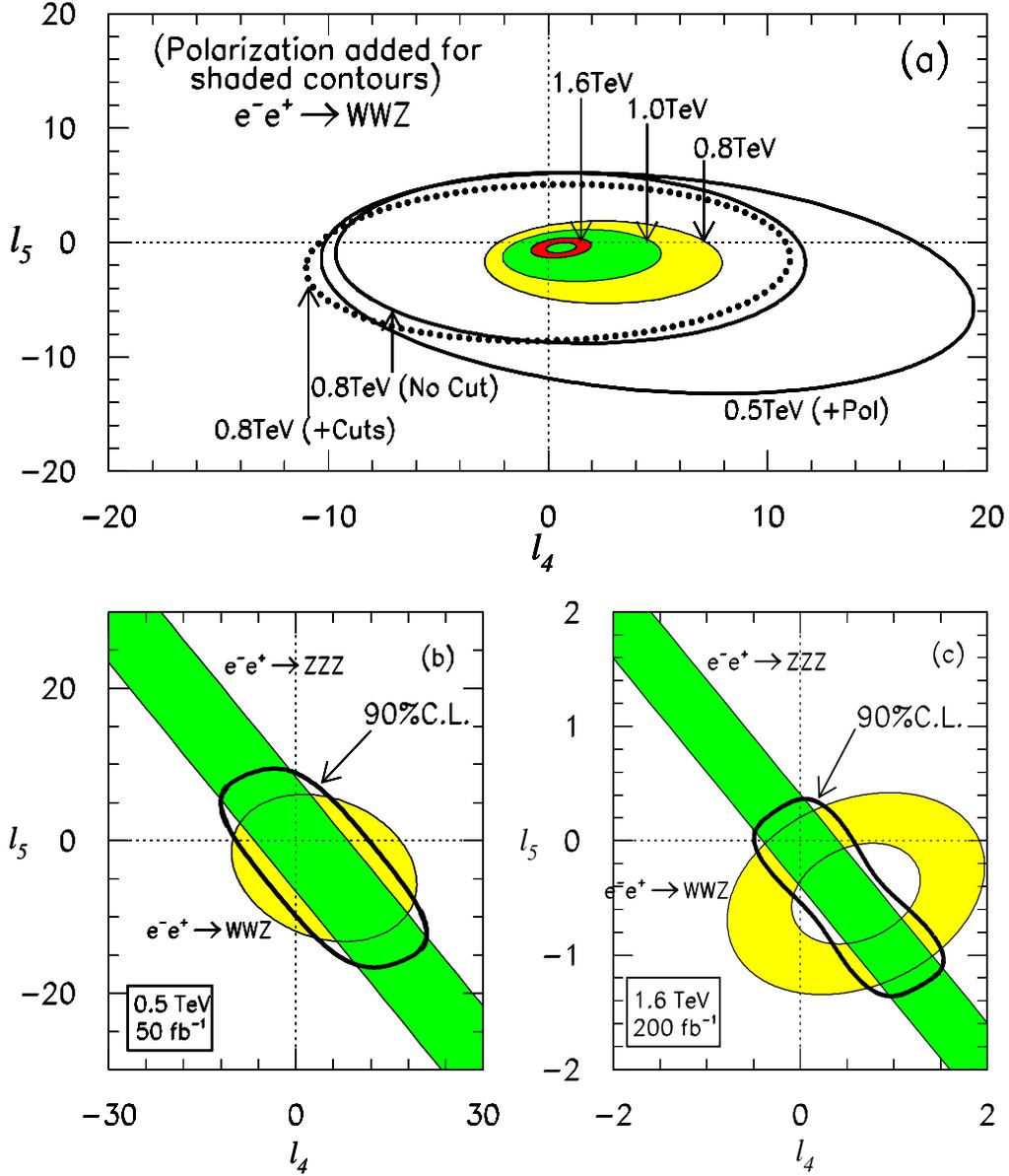,height=19cm,width=17cm}}
\vspace{-0.8cm}
\caption{Probing $\ell_4$-$\ell_5$ via $WWZ$ and $ZZZ$ production processes.
The roles of the polarization and the higher collider energy for
$e^-e^+\to WWZ$ are shown by the $\pm\,1\sigma$ exclusion contours  
in (a).
The integrated luminosities used here are 50~fb$^{-1}$ (at 500~GeV),
100~fb$^{-1}$ (at 800~GeV) and 200~fb$^{-1}$ (at 1.0 and 1.6~TeV).
In (b) and (c), the $\pm\,1\sigma$ contours are displayed for $ZZZ/WWZ$
final states at $\sqrt{s}=$0.5 and 1.6~TeV respectively,
with two beam polarizations ($90\% ~e^-$ and $65\% ~e^+$);
the thick solid lines present the combined bounds at $90\%$~C.L.
}
\label{fig1}
\end{figure}

%
\null   
\begin{figure}[T]       
\vspace*{-0.85cm}
\centerline{\epsfig{file=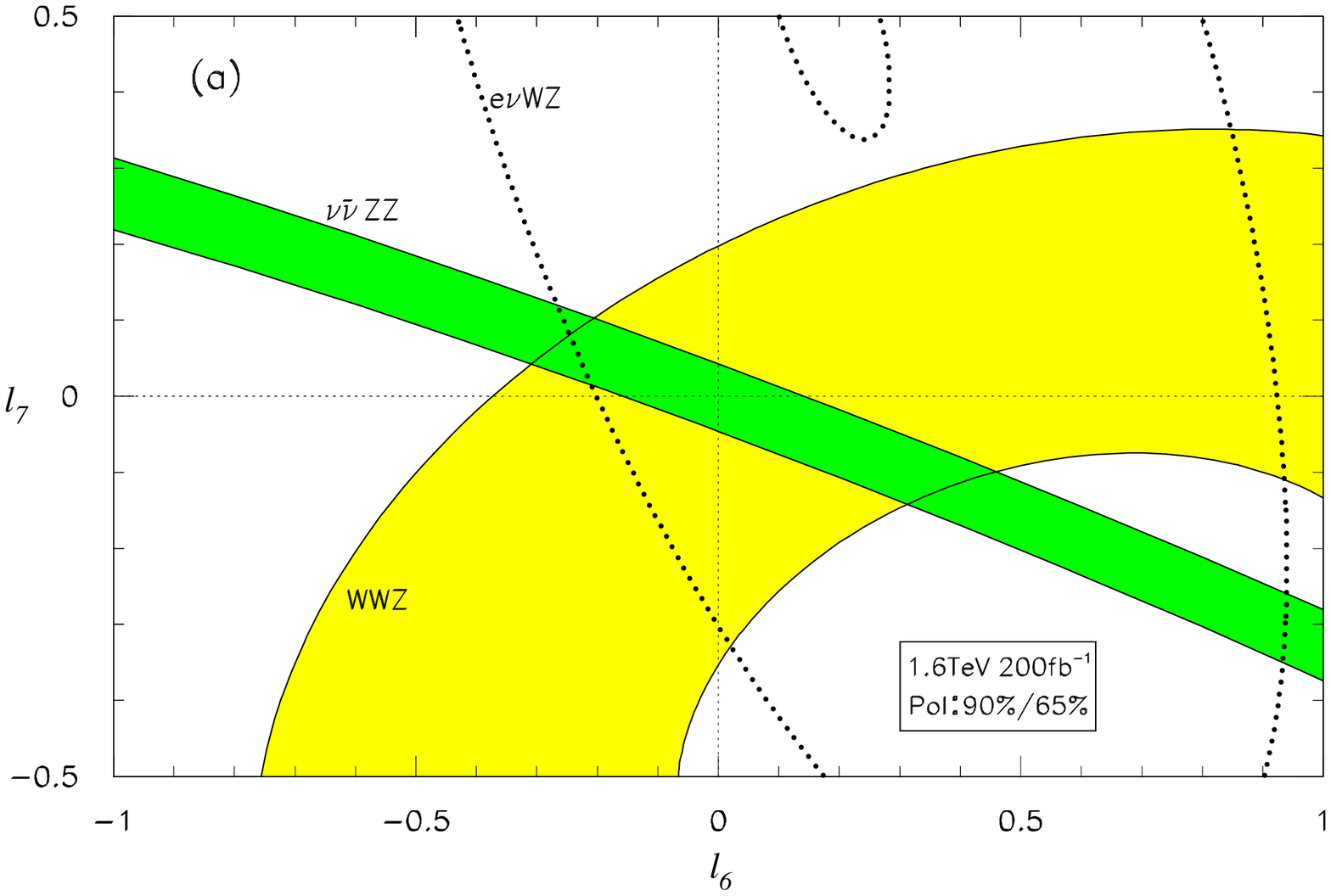,height=12cm,width=16cm}}
\vspace*{-1.35cm}
\hspace*{-0.2cm}
\centerline{\epsfig{file=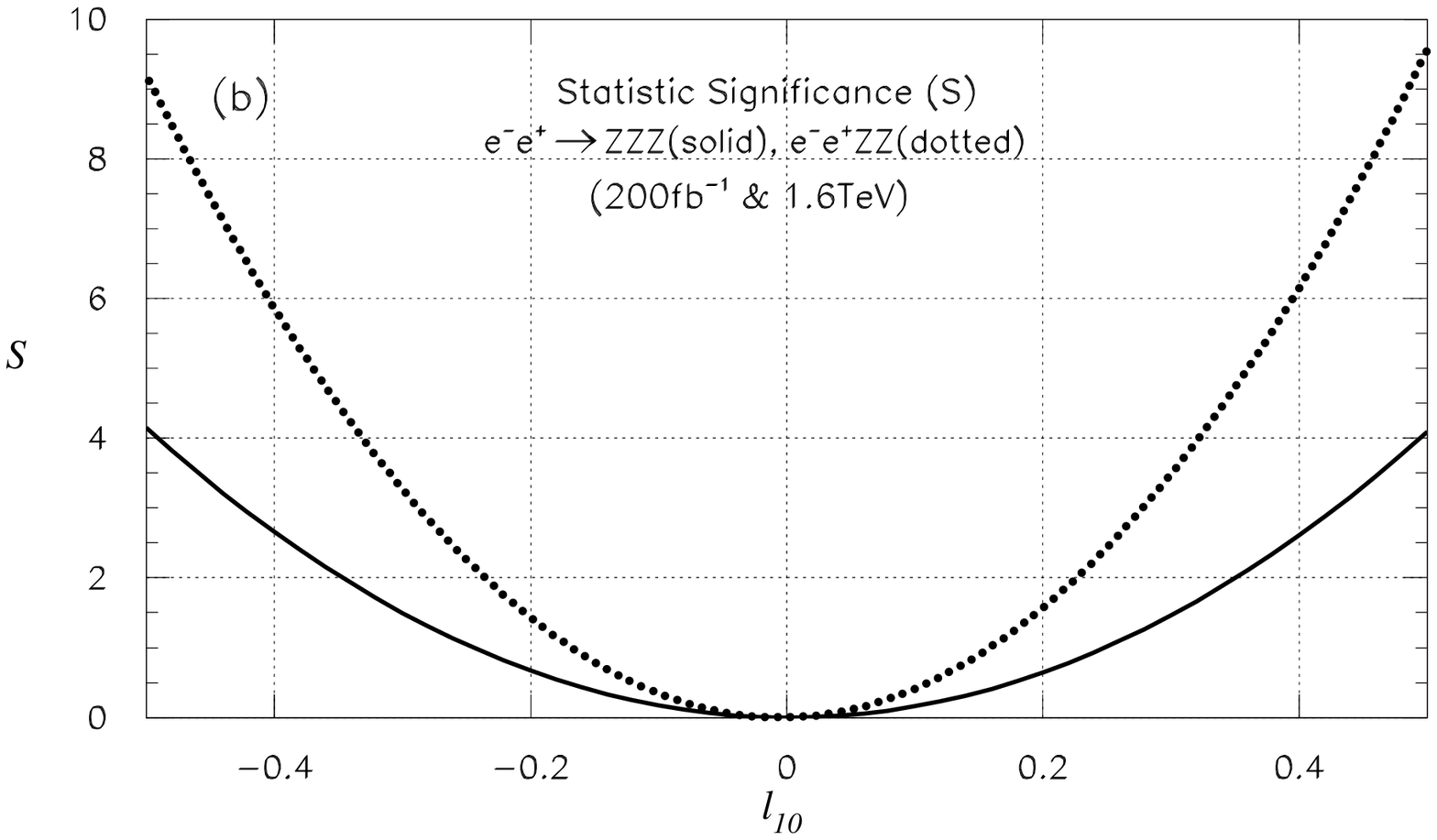,height=9.5cm,width=16cm}}
\vspace*{-0.25cm}
\caption{Interplay of the $WW$-fusion and $WWZ/ZZZ$-production for
discriminating the $SU(2)_c$-breaking parameters
$\ell_{6}$-$\ell_7$ and $\ell_{10}$
at $~\sqrt{s}=$1.6~TeV with $\int{\cal L}=$200~fb$^{-1}$:
(a). $\pm\,1\sigma$ exclusion contours for  $~e^-e^+ \to  
\nu\bar{\nu}ZZ,~
e^+\nu W^-Z/e^-\bar{\nu}W^+Z$,~ and $e^-e^+\to WWZ$~
with polarizations ($90\% ~e^-$ and $65\% ~e^+$).
(b). Statistic significance versus $\ell_{10}$
for $e^-e^+\to ZZZ,~e^-e^+ZZ$ (with unpolarized $e^\mp$ beams).
}
\label{fig4}
\end{figure}
\clearpage

\end{document}
\end

\end{document}
\end